\documentclass[
    ,final            
  ]
  {aipproc}

\layoutstyle{6x9}

\begin{document}

\title{Frequency Windows of Absolute Negative Conductance in Josephson 
Junctions}


\author{L. Machura}{
  address={Institute of Physics, University of Silesia, 40-007 Katowice, Poland}
}

\author{M. Kostur}{
  address={Institute of Physics, University of Augsburg, D-86135 Augsburg,
  Germany}
  ,altaddress={Institute of Physics, University of Silesia, 40-007 Katowice, Poland} 
}

\author{P. Talkner}{
  address={Institute of Physics, University of Augsburg, D-86135 Augsburg,
  Germany}
}

\author{P. H\"anggi}{
  address={Institute of Physics, University of Augsburg, D-86135 Augsburg,
  Germany}
}

\author{J. \L uczka}{
  address={Institute of Physics, University of Silesia, 40-007 Katowice, Poland}
}

\begin{abstract}
  We report on anomalous conductance in a resistively and capacitively
  shunted Josephson junction which is simultaneously driven by ac and  dc
  currents.  The dependence of the voltage across the junction on the
  frequency of the ac current shows windows of
  absolute negative conductance regimes, i.e. for a positive
  (negative) dc
  current, the voltage is negative (positive).
\end{abstract}

\maketitle

\section{Introduction}
A system in thermodynamic equilibrium responds to small
external stimuli in the way predicted by linear response theory. For
instance the current in an ohmic resistor increases with increasing
voltage. This intuitive situation where the effect follows the cause may 
change when the system is far from equilibrium. In the case of an 
electic circuit, it can happen that the increase of the voltage 
diminishes the current  or even induces a current in the opposite
direction so that
the conductance becomes negative. Such phenomenon is usually referred
to as an  'anomalus' response, in contrast to the 'normal', ohmic-like
response. There exists a broad variety of physical systems which can
exhibit such 'anomalous' behavior.
One of them is the so-called ratchet effect \cite{zapata}. 
Next is the negative differential mobility \cite{NDM,KosMac2006c}
of a massive particle moving in spatially periodic structures, driven
by a symmetric unbiased time-periodic force and thermal equilibrium
fluctuations \cite{LucEPL}. Further examples are absolute negative
conductance (ANC) or mobility (ANM) which were experimentally
confirmed in p-modulation-doped multiple quantum-well structures
\cite{hop} and semiconductor superlattices \cite{keay}. ANC (ANM) was
also studied theoretically for ac-dc-driven tunnelling transport
\cite{HarGri1997}, in the dynamics of cooperative Brownian motors
\cite{BroBen2000}, for Brownian transport in systems of a complex
topology \cite{EicRei2002a} and in some stylized,
multi-state models with state-dependent noise \cite{CleBro2002}, to
name but a few.  Recently, ANC was discovered in relatively
simple symmetric periodic systems \cite{MacKos2007a} like a Josephson
junction. In this paper, we continue to study the same system as in
\cite{MacKos2007a}, and demonstrate multiple manifestations of ANC in
the range of ac-current frequency.

\section{Stewart-McCumber model }
The Stewart-McCumber model of a Josephson junction is well known in
literature \cite{kautz}. In this model, a current through the junction
consists of a Josephson supercurrent characterized by the critical
current $I_0$, a normal current characterized by the resistance $R$
and a displacement current accompanied with the capacitance $C$.
Thermal equilibrium noise is the Johnson noise associated with the
resistance $R$. The dynamics of the phase difference $\phi=\phi(t)$ across
the junction is described by the following equation \cite{kautz}
\begin{eqnarray} \label{JJ1}
\Big( \frac{\hbar}{2e} \Big)^2 C\:\ddot{\phi} + \Big( \frac{\hbar}{2e} \Big)^2 \frac{1}{R} \dot{\phi}
= - \frac{\hbar}{2e} I_0 \sin (\phi) + \frac{\hbar}{2e} I_d + \frac{\hbar}{2e}I_a \cos(2 \pi \nu t) +
\frac{\hbar}{2e} \sqrt{\frac{2 k_B T}{R}} \:\xi (t),
\end{eqnarray}
where the dot denotes differentiation with respect to time $t$, $I_d$
and $I_a$ are the amplitudes of the applied dc and ac currents, respectively,
 $2\pi
\nu$ is the angular frequency of the ac driving.  The parameter $k_B$
denotes the Boltzmann constant and $T$ stands for temperature of the
system. Thermal equilibrium fluctuations are modeled by
$\delta$-correlated Gaussian white noise $\xi(t)$ of zero mean and
unit intensity.

The dimensionless form of the above equation reads
\begin{equation}
\label{JJ2}
\ddot{x} + {\gamma} \dot{ x} =- \sin (2\pi x) + F_0 + a \cos(\omega s) 
+ \sqrt{2 \gamma D_0} \; \Gamma(s),
\end{equation}
where $x=\phi/2\pi$, the dot denotes differentiation with respect to
the dimensionless time 
$s=t/\tau_0$ %
and 
$\tau_0=2\pi \sqrt{(\hbar/2e)  C / I_0}$ \cite{MachuraJPC}.  
The remaining re-scaled parameters become: the friction coefficient
is ${\gamma} = \tau_0 / RC$, the amplitude and the angular frequency of
the alternating current are denoted by   $a = 2\pi I_a / I_0$ and 
$\omega = 2 \pi \nu \tau_0$, respectively.  The rescaled bias load
reads $F_0= 2\pi I_d/ I_0$, the rescaled zero-mean Gaussian white
noise $\Gamma(s)$ has the auto-correlation function %
$\langle \Gamma(s)\Gamma(u)\rangle=\delta(s-u)$, and the noise
intensity $D_0 = (2e/\hbar)\: k_B T / I_0$.  The rescaled stationary
voltage
\begin{equation}
\label{V}
V= \langle \dot{ x} \rangle, 
\end{equation}
where the brackets $ \langle \dots \rangle $ denote an average over
the initial conditions, over all realizations of the thermal noise and
over one cycle of the external ac driving.  The dimensional voltage reads 
${\cal{V}}=(\hbar/2e)\omega_0 V $, where   
$\omega_0=(2/\hbar)\sqrt{E_JE_C}$
is the Josephson plasma frequency expressed by the Josephson coupling
energy $E_J=(\hbar/2e)I_0$ and the charging energy $E_C=e^2/C$.

Eq. (\ref{JJ2}), in the classical mechanics terms, describes a
Brownian particle moving in the spatially periodic potential
$U(x)=U(x+L)= -\cos(2\pi x)$ of unit period $L=1$, driven by the
time-periodic force and the constant force $F_0$.

\section{Absolute  negative conductance}
The noiseless deterministic dynamics determined by (\ref{JJ2}) shows 
rich behavior in dependence of the system's parameters and often 
exhibits  chaotic dynamics \cite{kautz}.  By
adding noise, the dynamics becomes diffusive i.e.  stochastic escape
events among possibly coexisting attractors are possible.  It follows
from the symmetry properties that the voltage is zero if the dc
current is zero ($F_0=0$). If the dc current is applied ($F_0\ne 0)$,
the reflection symmetry $x \to -x$ is broken and the voltage can
be non-zero. Most often, the signs of current and voltage coincide. 
Its sign is usually the same as the sign of $F_0$.
Sometimes, however, the dynamics turns out to be very counterintuitive
and regimes of ANC occur.  

The occurence of ANC may be governed by different mechanisms. 
In some regiemes ANC is induced by thermal equilibrium fluctuations, i.e.  
the effect is absent in absence of thermal fluctuations.  In other regimes, 
ANC can occur in the noiseless, deterministic system and the 
effect disapears with increasing temperature.
\cite{MacKos2007a}.  Both
situations though have its origins in the deterministic structure 
of the dynamics governed by stable and unstable orbits.

Here we report on a search for an optimal robust set of parameters   
where slight changes
of the system's parameters do not destroy the effect of ANC.  Based
on the original set of parameters ($a=4.2$, $\gamma=0.9$, $F_0=0.1$,
see \cite{MacKos2007a} for details) we scrutinize the interesting 
range of frequencies of the ac current.  
We have carried out extensive numerical simulations in
order to find as large as possible regimes of  pronounced ANC.  
Though  simulations also are 
time-consuming, they still require much less effort than performing 
a real experiment involving Josephson junctions.
\begin{figure}\label{fig1}
  \centerline{\includegraphics[width=0.85\textwidth]{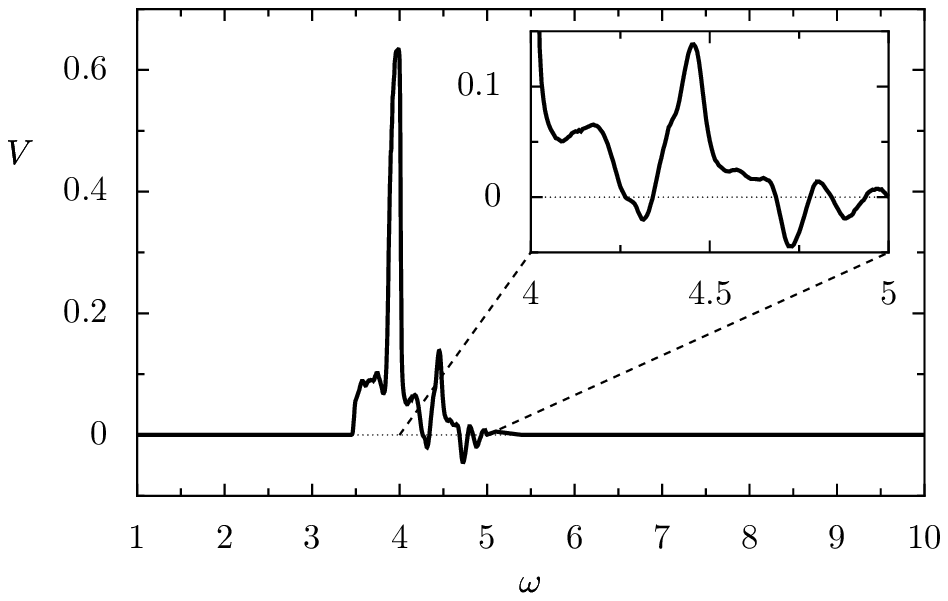}}
  \caption{ %
    The rescaled voltage $V$ versus rescaled angular frequency
    $\omega$ of the applied ac current.  Other parameters
    are set as following: $a=4.2$, $\gamma=0.9$, $D_0=0.001$ and the
    positive dc current $F_0=0.1$. Remarkable are  three regimes of
    frequencies where ANC occurs (see inset).}
\end{figure}

In Figure \ref{fig1} we present the average dimensionless voltage $V$
versus frequency of the external, time-periodic current. Within the
displayed range of frequencies one can distinguish three interesting
regions where ANC occurs. Up to frequency around $\omega=3.3$ the
average current is zero valued (numerically the current is never zero,
but it is smaller than a line thickness).  Starting from the mentioned
value the transport sets in and later the voltage reaches a
maximum of $V \approx 0.6337$ at  $\omega \approx 3.985$.
For higher frequencies, the voltage stays positive up to the value of %
$\omega \approx 4.260$ and then changes sign for the first time,
resulting in an anomalous response - the voltage becomes
\emph{negative} although the external dc current is in fact
\emph{positive}. This situation holds up to $\omega \approx 4.34$ and
then the voltage turns to positive values again. The same
scenario is repeated twice -- between the values of %
$\omega \approx 4.69 \div 4.79$ and $4.84 \div 4.94$. It means that
the phenomenon of ANC is not very rare and therefore gives more than
one possibility of adjusting  feasible frequencies of the  applied ac current 
 in  real experiments involving a single Josephson junction.
We also checked the system response for the signal
of other frequencies from the interval $\omega \in [10^{-4},10^2]$.  
However, we have not detected the ANC appearance.

In conclusion, we have shown that there are frequency windows of ANC
in a resistively and capacitively shunted Josephson junction driven by
ac current and constant dc current.  This phenomenon is quite robust
against variation of the system parameters which is desirable from the
experimental point of view.

\begin{theacknowledgments}
The work supported in part by the MNiSzW Grant N 202 131 32/3786.
\end{theacknowledgments}

\bibliographystyle{aipprocl}

\end{document}